\documentclass{revtex4}

\usepackage{epsfig}

\begin{document}

\title{The flow rate of granular materials through an orifice}

\author{ C. Mankoc , A. Janda , R. Ar\'evalo , J.M. Pastor , I. Zuriguel , A. Garcimart\'{\i}n and D. Maza }

\affiliation{ Departamento de F\'{\i}sica y Matem\'atica Aplicada \\
Facultad de Ciencias, Universidad de Navarra\\
31080 Pamplona, Spain\\
}

\begin{abstract}

The flow rate of grains through large orifices is known to be
dependent on its diameter to a $5/2$ power law. This relationship
has been checked for big outlet sizes, whereas an empirical fitting
parameter is needed to reproduce the behavior for small openings.

In this work, we provide experimental data and numerical simulations
covering a wide span of outlet sizes, both in three and two
dimensions. This allows us to show that the laws that are usually
employed are satisfactory only if a small range of openings is
considered. We propose a new law for the mass flow rate of grains
that correctly reproduces the data for all the orifice sizes,
including the behaviors for very large and very small outlet sizes.

\keywords{Granular flow \and Silos \and Jamming}
\end{abstract}

\maketitle

\section{Introduction}

The flow of granular materials through an orifice has been widely
studied during decades due to the great interest for industrial
applications ranging from silos to hoppers
\cite{Nedderman1,Tuzun,Savage,Tighe}. Contrary to the fluids, when a
silo is discharged by gravity, the flow rate does not depend on the
height of the granular layer. Indeed, when the thickness of the
layer is greater than a value close to $1.2$ times the diameter of
the silo, the pressure at the bottom of the silo saturates due to
the Janssen effect and hence, the flow rate remains about constant.
In a first order approximation it has also been shown that the flow
rate is independent on the diameter of the silo $L$ if two
conditions are fulfilled: $L$ is greater than 2.5 times the diameter
of the outlet orifice $D_0$ and also greater than $D_0+30\;d_p$,
where $d_p$ is the diameter of the particle \cite{Nedderman1}.

The most widely accepted law that predicts the flow rate of grains
through an orifice and its dependence on different parameters was
proposed by Beverloo et al. \cite{Beverloo1} and has the form:

\begin{equation}
\label{eqBeverloooriginal}
W=C \rho_b \sqrt{g}(D_0-kd_p)^{5/2}
\end{equation}
where $W$ is the average mass discharge rate through the orifice,
$C$ and $k$ are empirical discharge and shape coefficients
respectively, $\rho_b$ is the apparent density, and $g$ is the
acceleration of gravity. As $C$ may depend on the friction
coefficient, sometimes this is explicitly stated by writing
$C(\mu)$. Equation (\ref{eqBeverloooriginal}) is known as the
Beverloo law and its validity has been tested for mono-sized
granular samples with $d_p$ larger than 0.5 mm and $D_0$ big enough
to avoid intermittencies in the flow due to jamming. This means that
the flow rate of grains through orifices has been found to follow
the Beverloo law only for $D_0 \gg d_p$, well beyond the critical
value below which the flow can be interrupted due to the formation
of arches or domes \cite{Zuriguel1}.

One of the most interesting issues concerning Eq.
(\ref{eqBeverloooriginal}) is the dependence of the flow rate with a
$5/2$ power of the diameter of the orifice. This relationship can be
obtained by dimensional analysis and can be physically explained if
it is assumed that the granular flow is driven by the behavior of
grains near the outlet. Following this line of reasoning, it seems
plausible to believe that, just above the outlet, there is a
free-fall zone limited by an arch. Above the arch the grains are
well packed and their velocities are negligible, whereas below the
arch the particles accelerate freely under the influence of the
gravity. If the characteristic size of this arch is somehow
proportional to the radius of the orifice, the velocity of the
grains through the outlet of the silo is the one corresponding to a
particle falling without initial velocity from a distance
proportional to the radius of the outlet. A primary consequence of
this assumption is that the velocity of the grains is proportional
to $D_0^{1/2}$ and therefore the flow rate proportional to
$D_0^{5/2}$. An equivalent equation to Eq.
(\ref{eqBeverloooriginal}) can be easily derived for the two
dimensional case and a flow dependence on the diameter of the silo
of $D_0^{3/2}$ is obtained.

The Beverloo law, and particularly the $5/2$ power dependence of the
flow rate on the diameter of the orifice, has been found to be
robust for different kind of particles, independently of their
packing fraction, density, surface properties or shape
\cite{Hirshfeld,Hirshfeld2} for $D_0 \gg d_p$. It is also remarkable
that this equation can describe the flow rate of grains through
orifices for different flow patterns developed inside the silo, i.e.
massic flow, funnel flow or mixed flow. A modification has been
implemented for the flow of powders when the size of the particle is
smaller than 0.5 mm. In this case a term should be included in
equation (\ref{eqBeverloooriginal}) to reproduce the effect of the
pressure gradient generated by the air passing through the
interstices between the grains. Other modifications have been
developed to predict the solid discharge rate for binary mixtures
\cite{Tuzun,Humby} and even the dense flow of air bubbles in a two
dimensional silo \cite{Vandewalle}.

Despite the above mentioned robustness of the Beverloo law and the
fact that has been successfully used by engineers since 1961, the
basic physical principles behind the flow of grains through an
orifice remain elusive and the two empirical coefficients ($C$ and
$k$) are required to be determined experimentally for every single
kind of grains and container properties. The value of $C$, the
so-called discharge coefficient, depends on the bulk density and it
was found to be in a range of $0.55<C<0.65$ by Beverloo et al
\cite{Beverloo1}. The shape coefficient $k$ is generally agreed to
be dependent on the particle shape as well as the slope of the
hopper. However the meaning and origin of the term $-kd_p$ has
provoked a great controversy. The first and most widely accepted
interpretation of this term was done by Brown and Richards
\cite{BrownandRichards} who claim that the centers of the particles
cannot approach the edge of the orifice within a distance of
$kd_p/2$. Therefore the particle centres must pass through a
effective orifice of diameter $D_0-kd_p$. The value of $k$ has been
found to be independent of the size of the particle
\cite{Nedderman2} in a range of $1<k<2$ depending on the particle
and hopper properties. Yet there are some exceptions, like the flow
of sand, where the value of $k$ turns to be $2.9$.  However, Zhang
and Rudolph \cite{Zhang} claim that the only plausible value for $k$
is 1 and propose an alternative expression where a new term
$c_{\tau}$ is introduced. The value of $c_{\tau}$ depends on $D_0$
and physically represents the effect on the flow rate of shear
friction between flowing and non-flowing particles at the edges of
the orifice.

The experimental and numerical studies where the flow rate has been
found to be in a reasonably good agreement with the Beverloo fit are
abundant in the literature. Yet it is difficult to find any work
where the flow rate is measured covering, at least, two decades of
outlet sizes. Some of the works deal with very big orifices in real
silos where the value of $k$ has only a weak influence in the fit,
whereas others explore the region of small orifices but do not reach
high values of $D_0$ \cite{Hirshfeld2}. This fact may lead to a
misinterpretation of the results as the fit may seem suitable just
because the range of $D_0$ considered is not large enough and the
values of $C$ and $k$ are chosen arbitrarily. In this work we will
show that if the flow rate is measured for a wide range of orifice
sizes, the equation (\ref{eqBeverloooriginal}) is not able to
describe the whole behavior as different values of $C$ and $k$ must
be used for big and small orifices. In order to solve this problem a
new expression for the flow rate is proposed where the constant $k$
is altogether eliminated. The new equation is able to fit the
experimental results of the flow rate for diameters of the orifice
ranging from $1.5$ to $100$ times the diameter of the particle.
Furthermore with this modification it is also reproduced the
behavior of the flow rate for very small orifices, when the flow is
not continuous due to arch formation. Its validity has been checked
using different kind of particles in three and two dimensional
experiments as well as in two dimensional molecular dynamics
simulations.

The manuscript is organized as follows. In the first section a
description of the experimental setup and simulation techniques is
presented. The method used to measure the flow of grains is
discussed for both big and small orifices. In the next section the
numerical results in two dimensions and the experimental results in
two and three dimensions are reported. Different fits of Eq.
(\ref{eqBeverloooriginal}) will be displayed to stress that it is
unable to fit the flow rate for the whole range of outlet sizes.
Then, after measuring the deviations from the Beverloo proposal, we
introduce a modification consisting on the elimination of the
parameter $k$ and the correction of $W$ by a multiplicative term.
Finally we show that this corrective factor may have its physical
origin in a dependence of either the velocity or the apparent
density near the outlet of the silo.

\section{Methods and materials}

In order to investigate the flow through a small orifice under the
action of gravity, two scaled silos have been built. One of them is
cylindrical, and the other is two-dimensional (the beads are
contained between two glass panes so as to provide direct optical
access to the particles inside). Numerical simulations using
molecular dynamics have also been performed. In the following
subsections a description for the procedures used in each case is
provided.

\subsection{Three dimensional silo}

The experimental setup for the three dimensional silo was described
in detail in a former article \cite{Zuriguel2}. It consists of an
scaled, automated cylindrical container with an orifice in the base.
The silo is made of stainless steel, but vessels made of glass were
also used whenever observation of the particles was required. The
size of the silo is such that its finite size can be neglected.
Moreover, the level of the granular matter contained in it was never
allowed to descend below a certain level (about twice the silo
diameter) to ensure that the pressure at the base was approximately
constant due to the Janssen effect, as explained above. The piece
forming the flat bottom of the silo is changeable. This allows us to
modify the size of the outlet, which ranges from $2$ to $50$ mm. The
granular material falls through the orifice and is collected in a
box placed on a scales. Eventually the particles can get jammed. Let
us call an \emph{avalanche} the ensemble of particles fallen between
two jamming events. The size of the avalanche is calculated by
dividing the weight of the avalanche by the weight of one bead. The
time that the particles have been flowing is measured by means of a
microphone that registers the noise made by the falling grains.
Thus, with our experimental setup we can obtain both the size of
avalanches and its time span. After each avalanche, the system is
unjammed by directing a compressed air jet to the silo outlet; with
this device, the compaction fraction of the grains inside the silo
is not altered.

We used spherical grains of different sizes and made of different
materials. A summary of some of their relevant properties is shown
in table \ref{tab:granos}.

\begin{table}[htp]
\begin{tabular}{|c|c|c|c|}
\hline
Material & $d_p$ (mm)       & mass (mg)      & $\rho$ $(g/cm^3)$ \\
\hline
glass   & 0,42 $\pm$ 0,05 & 0,16 $\pm$ 0,05& 2,2  $\pm$ 0,1   \\
glass   & 1,04 $\pm$ 0,01 & 1,3  $\pm$ 0,4 & 2,4  $\pm$ 0,1   \\
glass   & 2,06 $\pm$ 0,02 & 10,1 $\pm$ 0,3 & 2,2  $\pm$ 0,1   \\
glass   & 3,04 $\pm$ 0,02 & 34,7 $\pm$ 0,4 & 2,4  $\pm$ 0,1   \\
lead    & 1,98 $\pm$ 0,06 & 46,0 $\pm$ 3,8 & 11,4 $\pm$ 0,5   \\
lead    & 3,0  $\pm$ 0,1  & 150  $\pm$ 14  & 10,9 $\pm$ 0,5   \\
delrin  & 3,00 $\pm$ 0,02 & 18,9 $\pm$ 0,3 & 1,34 $\pm$ 0,05  \\
\hline
\end{tabular}
\caption{Characteristics of the different beads that have been used
in the cylindrical silo. All the beads are spherical.}
\label{tab:granos}
\end{table}

\begin{figure}[!htb]
\centering
\includegraphics[keepaspectratio=true, width=9cm ]{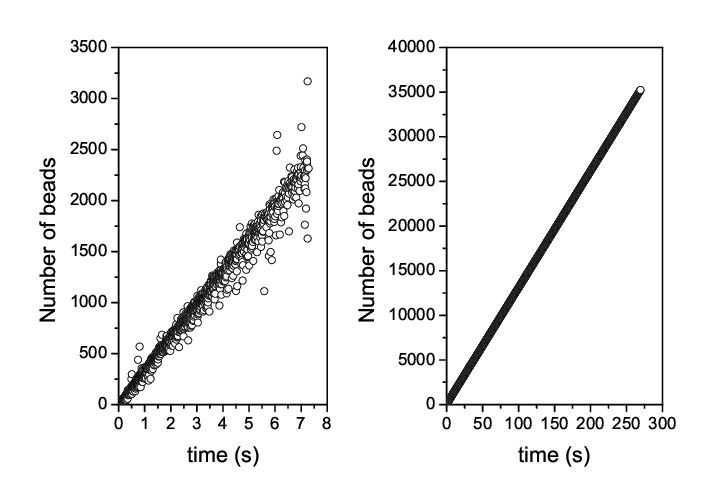}
\caption{Number of particles fallen as a function of time for a
small ($R=3.5$) and a big ($R=5.07$) outlet orifice, measured in the
three dimensional silo.} \label{fig:flujos}
\end{figure}

Flow measurement was accomplished in two different ways depending on
the features of the silo discharge. It has been shown that for big
sizes of the outlet orifice the outpouring is almost continuous,
while for a small orifice jamming events can occur. The border
between the two regimes is sharply defined at $R_c \approx 5$, where
$R$ is the ratio between the diameter of the outlet orifice and the
diameter of the beads \cite{Zuriguel2}. For $R>R_c$, the flow can be
measured just by dividing the number of fallen beads by the time
elapsed. As the flow is continuous, this measurement can be done at
any moment with any number of beads (or elapsed time), and it gives
consistent results. However, when $R<R_c$, the flow was
obtained by dividing the size of avalanches by their respective
duration. In order to illustrate the difference between both cases,
in Fig. \ref{fig:flujos} we show the measurements (number of fallen
beads as a function of time) for two outlet orifices, one bigger
than $R_c$ and other smaller.

In both cases, a measurement of the mean flow is obtained from the
slope of the straight line fitting the data. The main difference is
that for $R<R_c$ fluctuations are not completely smoothed out as in
the case $R>R_c$. This is not due to a lack of resolution in the
measurements but to the fact that at the involved time scales the
flow fluctuations can not be neglected for short avalanches. For
orifice sizes where both methods are feasible it has been checked
that the measured flow rate does not depend on the procedure.

\subsection{Two dimensional silo}

A two dimensional silo consisting of two sheets of glass was also
used. The silo is built so that the gap between the panes is a
little bit larger than the diameter of the beads. In order to meet
the stringent tolerances of these assembly, we only used stainless
steel beads with a diameter $1.00 \; \pm 0.01\; mm$, while the
separation between the two glass sheets was $1.10 \; mm$. As the
separation between walls is just a little bigger than the size of
the particles, they only can arrange themselves in one layer. The
setup therefore allows us to record the movement of individual
particles using a high speed camera. The grains flow out of the silo
through a slot in the base whose length can be changed at will. The
automation and measuring devices are similar to those used in the
three dimensional silo. A detailed description of this setup will be
presented elsewhere.

The time span of the avalanches is registered with a photodetector
at the silo exit. We placed optical fibers at both ends of the slit
forming the outlet orifice; a light beam is emitted from one of them
and the other collects it and feeds the photodetector, so a falling
bead is detected when it blocks the light beam. The time resolution
of this assembly (better than one millisecond) is smaller than the
time that it takes for a particle to cross the light beam. The
weight used also allows us to measure the avalanche size with a
resolution of just one particle.

The width of the silo is larger than $200$ particle diameters, in
order to prevent any influence of the lateral walls. As in the three
dimensional case, the measurement of the flow was accomplished using
two different methods, as described in the previous subsection,
depending on whether the outlet gets jammed or not.

\subsection{Numerical simulations}

We have used soft particle molecular dynamics \cite{Schafer}
simulations of disks in two dimensions. In this method, we consider
that a collision has taken place when two disks slightly overlap
(the distance between their centers is smaller than the sum of their
radii). A repulsive force, in the normal direction of the collision,
proportional to the overlap, prevents the grains from traverse each
other; besides, a term proportional to the relative velocity of the
particles accounts for the dissipation of energy during the
collision. A force perpendicular to the collision direction
implements the Coulomb law of friction. Details of the algorithm can
be found in \cite{Arevalo1}. The velocity-Verlet scheme along with
neighbor lists \cite{Rapaport} were used to integrate the equations
of motion.

A simulation begins with $5000$ disks placed on a regular lattice
that are given random velocities taken from a gaussian distribution.
The disks are allowed to fall under gravity through a hopper. Below
the hopper there is a flat bottomed silo in which the grains are
deposited. This is the preparation phase, which is aimed to break
the correlations that the initial regular arrangements of the grains
may induce in their dynamics. Once all the grains have settled in
the flat silo and after most of the kinetic energy is dissipated,
the outlet at the bottom is opened, allowing the grains to fall. The
flow is measured by counting the number of beads that go out of the
silo for each simulation step.

The walls of both the silo and the hopper used are constructed with
grains. The interaction grain-grain is the same than the interaction
grain-wall, but the latter keep their positions fixed. Thus the
walls are rough and give rise to dissipative collisions.

The flat silo is $50$ grains diameter wide and the level reached by
the grains is approximately twice that value when the silo is
filled. During a simulation, the level of the grains is kept
constant by reintroducing the exiting grains at the top of the silo.
These grains are placed as close to the free surface as possible and
with velocities similar to those of the exiting grains, so as not to
perturb the flow. These conditions allow us to neglect any effect
from the wall or filling method on the flow rate.

\section{Results}

\subsection{Validity of Beverloo's law}

\begin{figure}[!htb]
\centering
\includegraphics[keepaspectratio=true,width=9cm]{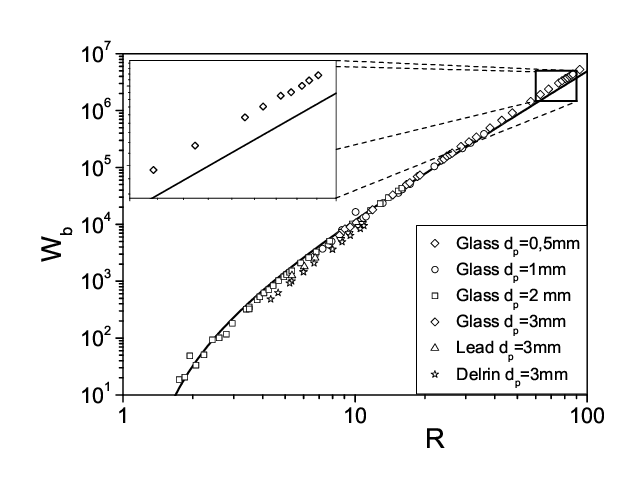}
\caption{Logarithmic plot of the flow rate of particles $W_b$ for a
three dimensional silo ($W_b$ is the number of beads fallen per unit
time). The solid line is the best fit of Eq.
(\ref{eqBeverloooriginal}): $W_b=50.5 (R-1.16)^{5/2}$.  (Note that
$g$ and $\rho$ are included in the numerical constant $C'=50.5$).
The material and the diameter of the beads corresponding to the
different symbols are provided in the legend. The inset shows an
expansion of the region from $R=60$ to $R=90$, where it can be seen
that $k=1.16$ does not produce a satisfactory match.}
\label{fig:flujo3Da}
\end{figure}

Let us start with the presentation of the flow measurements as a
function of $R$ in the three-dimensional silo. The experimental data
are shown on figure \ref{fig:flujo3Da}. As different materials are
included, we have taken the mass flow rate divided by the
mass of one bead $W_b=W/m_b$. By using $W_b$, which is the number of
beads fallen per unit time, all the data can be plotted together (as
expected, $W_b$ shows only a slight dependency on the material from
which the beads are made). Note that the size of the outlet orifice
spans over almost two decades, from $R=2$ to $R=100$. Besides, the
results are a practical demonstration that the relevant parameter is
indeed $R={D_0}/{d_p}$. The flow rate of beads made of different
materials can thus be represented in the same graph without any
additional concern.

The best fit obtained with Eq. (\ref{eqBeverloooriginal}) is
$W_b=50.5 (R-1.16)^{5/2}$ and is also represented in Fig.
\ref{fig:flujo3Da}. Note that this fit fails for large $R$, where it
underestimates the flow rate (see the inset in Fig.
\ref{fig:flujo3Da}). Indeed, the disagreement between the measured
and the predicted flow rate amounts to about one million beads per
second for $R=100$ (a relative error of 10\% approximately). Note
that the parameter $k$ merely shifts the fit along the horizontal
axis. Depending on the particular value chosen, the fit will be
closer to the data in a different zone of $R$. If the range of $R$
is small, this may seem acceptable; but if a large range of exit
orifices is considered, it becomes clear that Beverloo's law is
inadequate.

\begin{figure}[!htb]
\centering
\includegraphics[keepaspectratio=true,width=9cm]{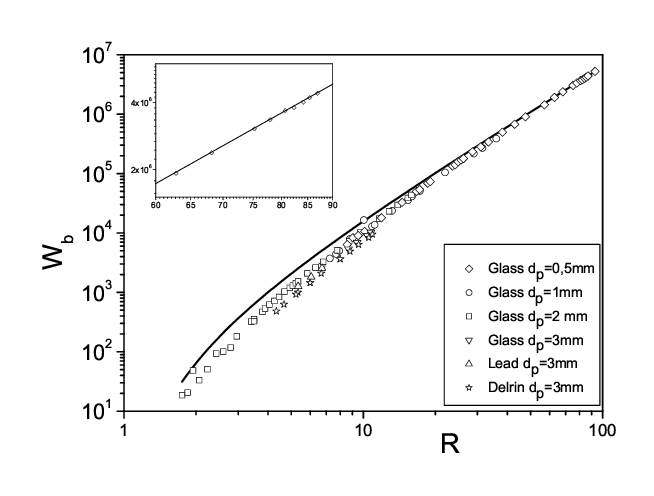}
\caption{Logarithmic plot of the flow rate of particles $W_b$ for a
three dimensional silo (same data than in Fig. \ref{fig:flujo3Da}).
The solid line is a fit of Eq.  (\ref{eqBeverloooriginal}) with
$k=1$: it gives $W_b=64.59 (R-1)^{5/2}$. \emph{Inset}: a zoom of the
same data and fit for the range $60<R<90$.}  \label{fig:flujo3D}
\end{figure}

A different approach that can be tried is to consider large values
of $R$ and fit Eq. (\ref{eqBeverloooriginal}) to the experimental
data with $k=1$. A different approach is to fit Eq.
(\ref{eqBeverloooriginal}) for large $R$ with $k=1$. In this case at
least the asymptotic behavior $W \rightarrow R^{5/2}$ for
$R\rightarrow\infty$ is recovered (see inset of Fig.
\ref{fig:flujo3D}). But then for small orifices this prediction and
the experimental data do not agree: the flow rate is overestimated
(see Fig. \ref{fig:flujo3D}).

\begin{figure}[!htb]
\centering
\includegraphics[keepaspectratio=true,width=9cm]{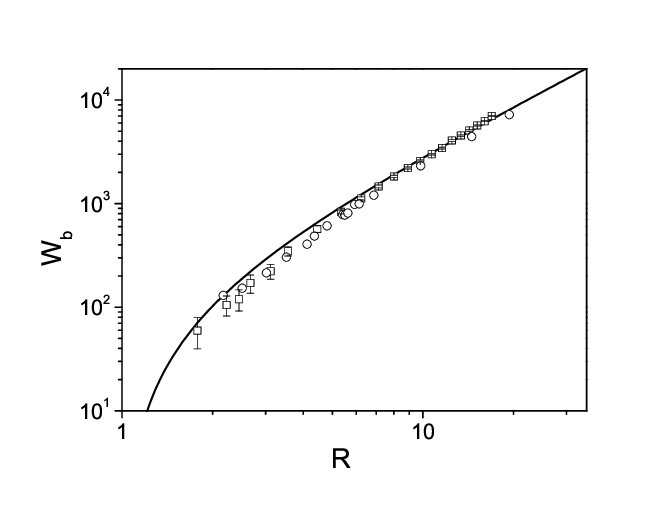}
\caption{Logarithmic plot of the flow rate of particles $W_b$ for a
two dimensional silo. Circles represent experimental measurements,
while squares are the result of numerical simulations. The error bar
for the experimental data is of the same order as the symbol size.
The solid line is a fit of Beverloo's law with $k=1$: $W_b=102.2
(R-1)^{3/2}$.} \label{fig:flujo2D}
\end{figure}

The same holds for the two-dimensional case, where the asymptotic
behavior for large $R$ is now $W_b \rightarrow R^{3/2}$ (Fig.
\ref{fig:flujo2D}). Note, however, that in this case we have
explored a smaller range of $R$. By varying the free parameters of
Eq. (\ref{eqBeverloooriginal}), \emph{i.e.} $C$ and $k$, one can
often get a seemingly acceptable result. Nevertheless, close
inspection of the fit with $k=1$ reveals the same qualitative
disagreement: Beverloo's law with $k=1$ overestimates the flow for
small $R$. This can be more easily seen if a large range of $R$ is
considered; but even for a smaller range of outlet sizes, such as in
in Fig. \ref{fig:flujo2D}, the overestimation is suggested for
$R\leq10$. In the same figure the result of the numerical simulation
is displayed. The agreement with the experimental results is
excellent for all the explored range.

In summary, the experimental results both in two and three
dimensions as well as the numerical simulations in two dimensions
clearly show a discrepancy with the values predicted by Eq.
(\ref{eqBeverloooriginal}). If a small range of large outlet
orifices is considered, the free parameters of Beverloo's equation
allow for a reasonably good fit. However if a large range of $R$ is
included, the fit does not yield acceptable results.

\subsection{Modification to Beverloo's statement}

\begin{figure}[!htb]
\centering
\includegraphics[keepaspectratio=true,width=9cm]{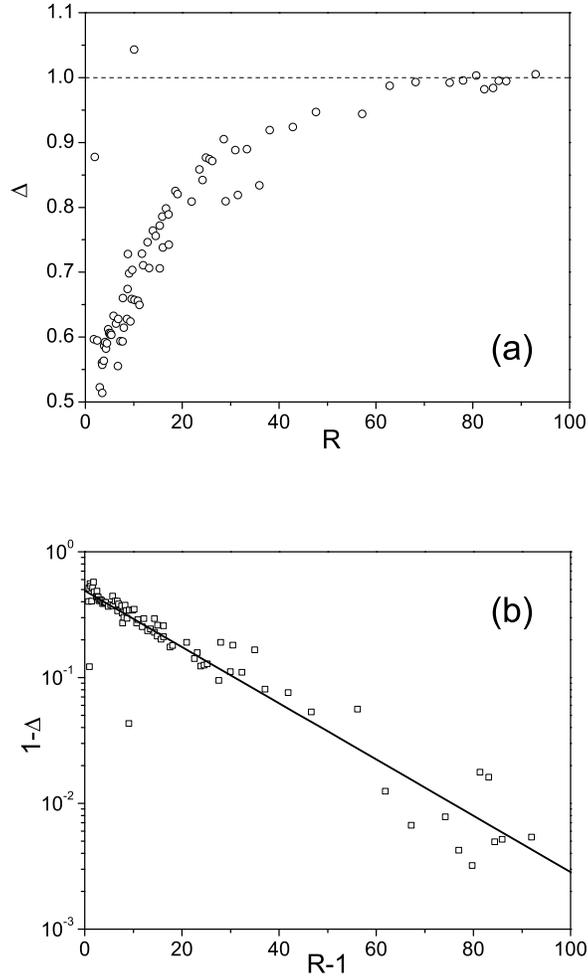}
\caption{(a) $\Delta$ versus $R$ for the three dimensional silo. (b)
Semilogarithmic plot of $1 - \Delta$ \emph{vs} $R-1$ and the
proposed fit with Eq. (\ref{eq:delta}), from which the value
$b=0.051$ has been obtained.} \label{fig:delta}
\end{figure}

In order to study more closely the discrepancy between the measured
data in the three dimensional case and the values predicted by the
Beverloo equation, we show in Fig. \ref{fig:delta}(a) $\Delta$
\emph{vs.} $R$, where $\Delta = W_m / W_p$ is the ratio between the
measured flow $W_m$ and the value $W_p$ predicted by Beverloo's law
with $k=1$, obtained from the fit shown in Fig. \ref{fig:flujo3D}.

It can be seen that $\Delta$ saturates in a non linear fashion. For
large $R$ indeed $\Delta \rightarrow 1$, \emph{i.e.} $W_p = W_m$, as
expected by construction. In order to check the exponential
dependence, we plot $1 - \Delta$ vs. $R-1$ on a semilogarithmic
scale (Fig. \ref{fig:delta}(b)). A fit is also provided showing the
good general agreement of this functional dependence:

\begin{equation}
\Delta = 1 - \frac{1}{2} e^{- b \cdot (R-1)} \label{eq:delta}
\end{equation}

From the fit, one can obtain that $b=0.051$.

We then propose another expression for the mass flow rate in number
of beads per unit time $W_b$, which stems from Eq.
(\ref{eqBeverloooriginal}) $W_p$ after the inclusion of the
correction factor $\Delta$:

\begin{equation}
W_b = C' \left ( 1 - \frac{1}{2} e^{- b \cdot (R-1)} \right )
(R-1)^{5/2} \label{eq:flu_mod}
\end{equation}

Recall that the constants $C$, $\rho_b$ and $\sqrt{g}$ from Eq.
(\ref{eqBeverloooriginal}) have been grouped into $C'$.

\begin{figure}[!htb]
\centering
\includegraphics[keepaspectratio=true,width=9cm]{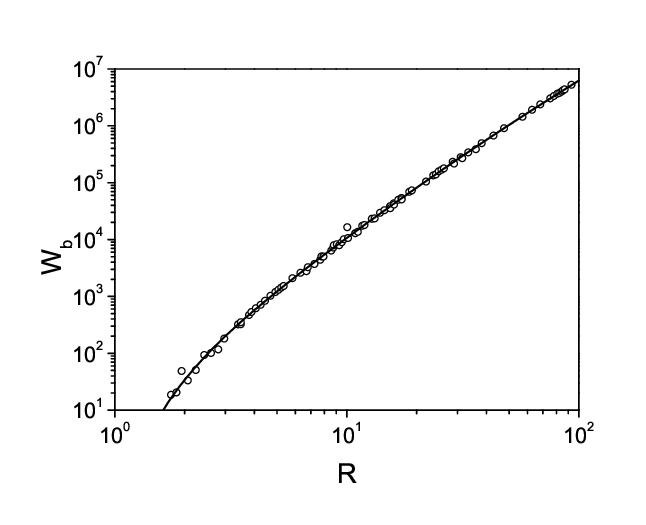}
\caption{Fit of the experimental data for the 3d silo using the
proposed equation for the flow rate (eq. (\ref{eq:flu_mod})) with
$C'= 64.6$ and $b=0.051$. The data are the same than in Fig.
\ref{fig:flujo3D}, taking only the ones corresponding to glass
beads.} \label{fig:ajuste3D}
\end{figure}

In Fig. \ref{fig:ajuste3D} we show the fit of equation
(\ref{eq:flu_mod}) to our experimental data. As can be seen, the
match is very good in all the range of $R$, which spans for almost
two decades. It should be stressed that the proposed law fits the
mass flow rate even for very small orifices (\emph{i.e.} $R<5$),
where the flow can be interrupted by jamming events.

\begin{figure}[!htb]
\centering
\includegraphics[keepaspectratio=true,width=9cm]{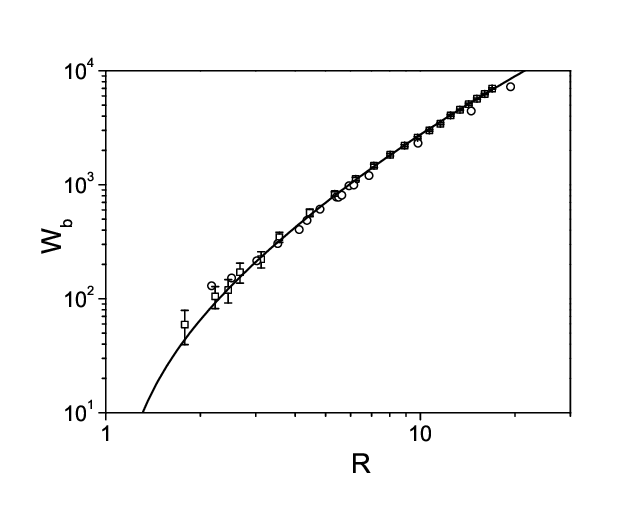}
\caption{Fit of the experimental data for the 2d silo (circles) and
the numerical simulations (squares) using the proposed equation for
the flow rate (eq. (\ref{eq:flu_mod})) with an exponent $3/2$. The
values of the parameters are $C'=108$ and $b=0.23$.}
\label{fig:ajuste2D}
\end{figure}

In two dimensions, the functional dependence is the same as in
three dimensions except for the exponent $5/2$ which is now $3/2$
(coefficients $C'$ and $b$ are obviously different). In figure
\ref{fig:ajuste2D} we show the experimental data and the results for
the numerical simulations along with the proposed fit.

\begin{figure}[!htb]
\centering
\includegraphics[keepaspectratio=true,width=9cm]{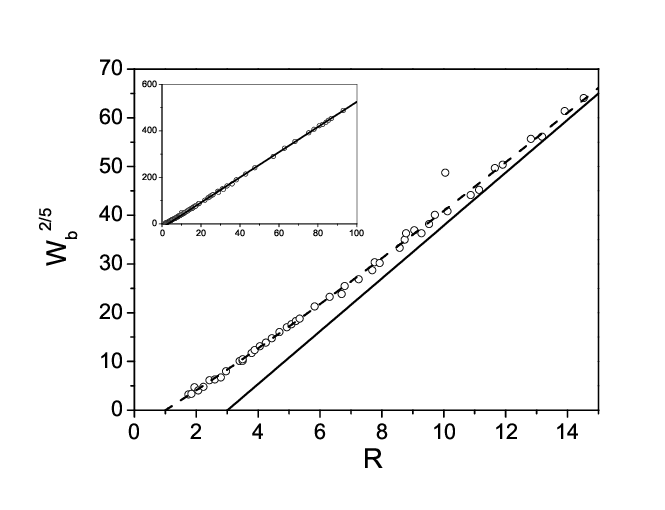}
\caption{ $W_b^{2/5}$ versus $R$ for the three dimensional silo near
$R=1$. The solid line corresponds to a linear fit for the data
corresponding to $R>50$ (the whole range of data is shown in the
inset). The dashed line is the fit using the proposed flow rate
equation (\ref{eq:flu_mod}). } \label{fig:W25}
\end{figure}

The proposed functional dependence of the flow is more satisfactory
than fitting the data by using the parameter $k$. Let us recall that
this parameter was introduced by Beverloo in order to fit the flow
for small orifices. The way that this is usually done is the
following. As $W \propto (D_0-k d_p)^{5/2}$ in Eq.
(\ref{eqBeverloooriginal}), then plotting $W^{2/5}$ vs. $R$ should
give a straight line and the intercept with zero would provide the
value for $k$. This is carried out, for example, in ref.
\cite{BrownandRichards2}. Our experimental data for the flow rate
when the orifice is small (Fig. \ref{fig:W25}) provide a strong
evidence that the intercept with zero is at, or very near to, $R=1$:
we have observed that grains flow until near that value. It should
be noticed that if the flow rate is not measured for small $R$ this
method can be misleading, as it depends on an unwarranted
extrapolation. Indeed, the value of $k$ obtained depends on the
range of $R$ used to perform the fit. For instance, the solid line
in Fig. \ref{fig:W25} is a linear fit obtained using just the points
for $R>50$, and the value obtained in this case is $k \approx 3$.
Any value between $k=1$ and $k \approx 3$ can be obtained with our
data simply by considering just a suitable range of $R$.

As explained above, the only plausible value for $k$ is unity. This
is reasonable since $W$ must vanish for $D_0\rightarrow d_p$, and so
one could just write $W\propto(D_0-d_p)$, with $k=1$. But then it is
impossible to fit correctly the data using Eq.
(\ref{eqBeverloooriginal}), as shown before (see Fig.
\ref{fig:flujo3D}). Yet the proposed equation (\ref{eq:flu_mod})
neatly fits the data for large values of $R$ and it gives the
correct behavior for $R\rightarrow 1$.

Furthermore, our data show that the so-called ``empty annulus''
effect is not evident at all. It is obvious that no bead can pass
through the orifice in such a way that its center is separated by
less than $d_p / 2$ from the orifice border. Thus, the bead centers
always pass through the orifice in a region which is effectively
determined by a diameter $D_0 - d_p$. If anything beyond this
obvious logic is considered, such as an ``empty annulus'' defined by
$D_0 - k d_p$ with $k \neq 1 $, then it should be checked that the
flow vanishes for $D_0 = k d_p$. We have found no hint of this;
instead we find $W_b> 0$ for values of $k$ often given in the
literature.

In our proposal, embodied in Eq. (\ref{eq:flu_mod}), we have got rid
of the parameter $k$ and we have instead introduced a correction
term, namely, the parenthesis containing the exponential. One of the
more appealing possibilities is that this correction term is related
to the density of grains near the outlet. In both the experiments
and the simulations the compaction fraction was the always the same
at the beginning of the runs, and we have checked that it remains
approximately unchanged during the discharge. But it may be
different near the outlet orifice. As this measurement is difficult
to obtain experimentally, we resort to numerical simulations in a
two dimensional silo. The value of the density depends on the size
of the region where numerical data are collected, and this should be
so because there exist spatial density variations near the outlet.
We show the results for the density in Fig. 9, where we plot the
local compaction fraction in a zone close to the outlet orifice as a
function of $R$. The measurements for two regions of different size
can be fit with a term just like the one proposed for the flow rate,
i.e. $\left ( 1 - \frac{1}{2} e^{- b \cdot (R-1)} \right )$. The
only variation between the fits is a multiplicative factor. We can
then conclude that the correction factor that we introduced is
related to the density variation of the granular material near the
outlet orifice.

\begin{figure}[!htb]
\centering
\includegraphics[keepaspectratio=true,width=9cm]{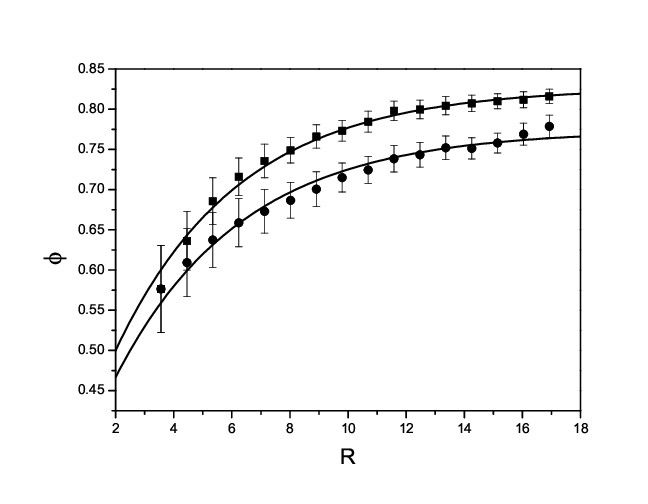}
\caption{The compaction fraction $\Phi$ as a function of $R$ for a
two dimensional silo, obtained from numerical simulations. The value
of the density depends on the size of the region where it is
measured: circles and squares correspond to two square regions just
over the outlet orifice. The fitting function in both cases is
$\left ( 1 - \frac{1}{2} e^{- b \cdot (R-1)} \right )$. }
\label{fig:dens}
\end{figure}

Finally, we show the robustness of the proposed expression by
displaying the residues of the best fit using both Eq.
(\ref{eqBeverloooriginal}) and Eq. (\ref{eq:flu_mod}). The fit given
by Beverloo is essentially flawed in the sense that it cannot
reproduce faithfully the flow for an extended range of $R$ (so the
residues are only small for one particular value of $R$). As
mentioned above, it is remarkable that for $R=100$ the difference
between the value given by Eq. (\ref{eqBeverloooriginal}) and the
measured value amounts to about one million beads per second. On the
contrary, the residues for Eq. (\ref{eq:flu_mod}) are consistently
small.

\begin{figure}[!htb]
\centering
\includegraphics[keepaspectratio=true,width=9cm]{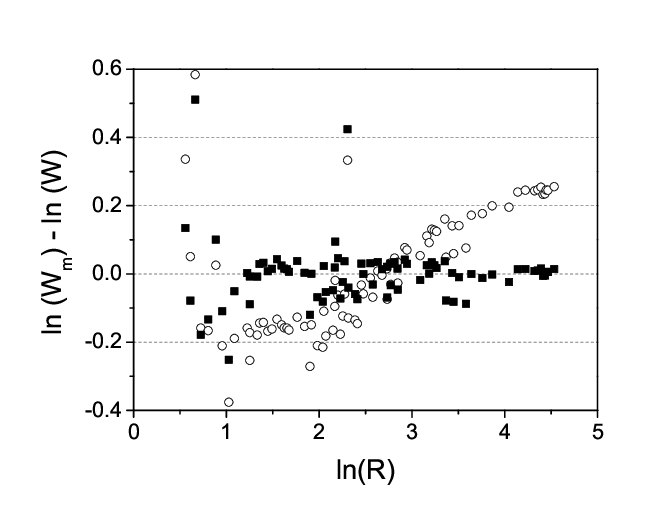}
\caption{The residues for the fits are shown as the difference
between the logarithms of the measured flow $W_m$ and the value $W$
provided by the best fit with the two equations considered.
\emph{Open circles}: the differences for the fit provided by
Beverloo (Eq. (\ref{eqBeverloooriginal}), with the parameters as
used in Fig. \ref{fig:flujo3Da}). \emph{Solid squares}: the
differences for the fit provided by Eq. (\ref{eq:flu_mod}) for the
the mass flow rate.} \label{fig:resid}
\end{figure}

\section{Conclusions}
In this paper we have looked into the behavior of the flow of
granular materials during the discharge of a silo. We have checked
that the flow $W$ does not depend significantly of the material from
which the beads are made, and that the relevant control parameter is
the ratio $R$ between the diameter of the orifice and the diameter
of the beads. We have collected experimental data for the largest
range of outlet sizes we have been able to find in the literature,
which comprises about two decades in $R$ and six decades in $W_b$.

Concerning the flow rate, the Beverloo expression provides the right
asymptotic behavior $W \rightarrow R^{5/2}$ for large $R$. However,
the Eq. (\ref{eqBeverloooriginal}) is not valid for a sizeable range
of $R$ including small sizes of the outlet orifice, \emph{i.e.}
$R\sim10$. In this case, the inclusion of the term $k d_p$ in the
flow equation, which was an arbitrary remedy for fitting the data,
cannot account for the experimental results. We propose instead a
more robust expression in which the flow depends on the product of
$R^{5/2}$ times an exponential corrective factor. We have also shown
that it fits neatly the data for a large range of $R$, including
very small orifices, in two and three dimensions.

In the new proposal for the mass flow rate, we have got rid of the
factor $k$ and introduced instead a correction term which saturates
exponentially with $R$, so that the asymptotic behaviour $R^{5/2}$
is recovered for large $R$. Apart from demonstrating that the
concept of ``empty annulus'' is wrong and unable to explain the
experimental results, we have provided an explanation for the
correction introduced in terms of the local density variations near
the outlet.

\begin{acknowledgements}
This work has been partially supported by Project FIS 2005-03881
(Spanish Government) and PIUNA (Universidad de Navarra). C.M.,
J.M.P. and R.A. thank Asociaci\'on de Amigos de la Universidad de
Navarra for a fellowship. A.J. thanks Fundaci\'on Ram\'on Areces for
a fellowship.
\end{acknowledgements}

\end{document}